\documentclass[a4paper,11pt]{article}
\usepackage{amsmath,amssymb,graphicx}
\usepackage{hyperref}

\usepackage{cite}

\usepackage{multicol,color,longtable}

\definecolor{darkred}{rgb}{0.65,0.15,0}
\hypersetup{pdfborder={0 0 0},colorlinks=true,urlcolor=darkred,citecolor=blue,linkcolor=darkred,linktocpage=true}

\newcommand{\doi}[2]{\href{http://dx.doi.org/#2}{#1}}

\newcommand{\eprintth}[1]{\href{http://arxiv.org/abs/#1}{#1}}
\newcommand{\eprintN}[2]{\href{http://arxiv.org/abs/#1}{arXiv:#1 [#2]}}

\newcommand{\e}{\mathfrak{e}_{10}}
\newcommand{\ke}{K(\mathfrak{e}_{10})}
\newcommand{\E}{E_{10}}
\newcommand{\KE}{K(E_{10})}

\newcommand{\ta}{{\mathtt a}}
\newcommand{\tb}{{\mathtt b}}

\newcommand{\mf}[1]{\mathfrak{#1}}
\newcommand{\nn}{\nonumber}

\begin{document}

\thispagestyle{empty}

\mbox{ }
\vspace{30mm}

\begin{center}
{\LARGE \bf Higher spin representations of $K(E_{10})$}\\[10mm]

\vspace{8mm}
\normalsize
{\large  Axel Kleinschmidt${}^{1,2}$ and Hermann Nicolai${}^1$}

\vspace{10mm}
${}^1${\it Max-Planck-Institut f\"{u}r Gravitationsphysik (Albert-Einstein-Institut)\\
Am M\"{u}hlenberg 1, DE-14476 Potsdam, Germany}
\vskip 1 em
${}^2${\it International Solvay Institutes\\
ULB-Campus Plaine CP231, BE-1050 Brussels, Belgium}

\vspace{20mm}

\hrule

\vspace{10mm}

\begin{tabular}{p{12cm}}
{\small
We review the recently constructed non-trivial fermionic representations of the 
infinite-dimensional subalgebra $\ke$ of the hyperbolic Kac--Moody algebra $\e$.
These representations are all unfaithful (and more specifically, of finite dimension).
In addition we present their decompositions under the various finite-dimensional subgroups 
associated with some maximal supergravities in dimensions $D\leq 11$, and the projectors for 
`spin-$\frac72$' which have not been given before. Those representations that
have not been derived from supergravity still have to find a role and a proper physical 
interpretation in the conjectured correspondence between $\E$ and M-theory. Nevertheless, they provide novel mathematical structures that could shed some light on fundamental questions in supergravity and on the possible role of $\KE$ as an `R-symmetry' of M-theory, 
and perhaps also on the algebra $\e$ itself.
}
\end{tabular}

\vspace{10mm}
\hrule

\end{center}

\newpage

\section{Introduction}

The hyperbolic Kac--Moody algebra $\e$ has been conjectured to generate an underlying symmetry of M-theory~\cite{Julia,Damour:2002cu}  and its (maximal compact) subalgebra $\ke$ (fixed by the Chevalley involution) plays the role of the generalised R-symmetry transformations~\cite{deBuyl:2005zy,Damour:2005zs,deBuyl:2005sch,Damour:2006xu,Kleinschmidt:2007zd}. In the $\e$ conjecture the constrained null motion of a spinning particle on the symmetric space $\E/\KE$ is equivalent to the dynamics of supergravity or even M-theory. This conjecture thus far has only been verified for a finite set of generators of the infinite-dimensional algebra $\e$ both in the bosonic and fermionic sector~\cite{Damour:2002cu,Kleinschmidt:2004rg,Kleinschmidt:2004dy,Damour:2005zs,deBuyl:2005sch,Damour:2007dt}. However, it has thus far proved impossible to construct a spinning particle action on $\E/\KE$ that has one-dimensional local supersymmetry, as was explained at length in Ref.~\cite{Kleinschmidt:2014uwa}.

One major obstacle when constructing a supersymmetric $\E$-model is the disparity between the bosonic and fermionic degrees of freedom that are used: The bosons are associated with the infinitely many directions of the symmetric space $\E/\KE$ whereas the fermions used in Refs.~\cite{Damour:2005zs,deBuyl:2005sch} were constructed out of a finite-dimensional 
(hence unfaithful) 
representation of dimension $320$ of the R-symmetry group $\KE$.\cite{const} It therefore appears necessary to construct larger, preferably infinite-dimensional, fermionic representations of $\KE$ and this is the topic we will pursue in the present contribution that is partially based on our paper Ref.~\cite{Kleinschmidt:2013eka}. 

We develop a new formalism for constructing representations of $\ke$ and exhibit new irreducible examples of dimensions $1728$ and $7040$, respectively. We refer to them as `higher spin representations' although their spin is not necessarily higher from a space-time point of view but rather when viewed from the (truncated)
Wheeler--DeWitt superspace of metrics. This point will be explained in more detail below. We will see that only the ${\bf 7040}$ contains also genuine higher spin fields from the space-time perspective.
Our formalism gives the action of an infinite number of $\ke$ generators that are labelled by the positive real roots of $\e$. Since the representations are finite-dimensional and therefore necessarily unfaithful, an infinite number of these generators will be represented by the same operator on the representation space.

Let us emphasize that a proper understanding of the fermionic sector will be
essential for further progress with understanding the role of $\E$ in M-theory, 
something that is unlikely in our opinion to be achievable if one restricts attention 
to the bosonic sector only. On top of
the (unknown) representation theory of $\KE$ this might quite possibly require some
novel type of bosonisation, as is suggested by the fact that $\E$ seems to `know everything'
about the fermions of maximal supergravity that we have learnt from supersymmetry
(in particular, the structure of the bosonic and fermionic multiplets). Equally important, the actual \textit{physics} of the quantised theory with fermions is likely to differ very much from that of the purely bosonic system, as is obvious from the example of supersymmetric quantum cosmology investigated in~Ref.~\cite{Damour:2014cba}.

\section{$\e$ and $\ke$}

\begin{figure}
\centerline{\includegraphics[width=8.5cm]{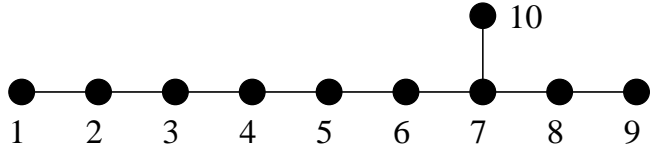}}
\caption{The Dynkin diagram of $\e$ with labelling of nodes.}
\label{fig:e10dynkin}
\end{figure}

The (split real) Lie algebra $\e$ is a hyperbolic Kac--Moody Lie algebra~\cite{Kac}. Its only known definition is in terms of generators and relations. There are $30$ generators $(e_i,f_i,h_i)$ for $i=1,\ldots,10$ and each triple generates an $\mf{sl}(2,\mathbb{R})$ subalgebra of $\e$. The full set of defining relations is given by
\begin{align}
[h_i, h_j] &=0,& 
[h_i, e_j] &= A_{ij} e_j,&
[h_i, f_j] &= -A_{ij} f_j,&\nn\\
{} [e_i,f_j] &= \delta_{ij} h_i,& 
(\mathrm{ad}\,e_i)^{1-A_{ij}} e_j &=0,&
(\mathrm{ad}\,f_i)^{1-A_{ij}} f_j &=0.
\end{align}
Here, $A_{ij}$ are the elements of the symmetric Cartan matrix associated with the $\e$ Dynkin diagram shown in figure~\ref{fig:e10dynkin}. The Cartan matrix is of Lorentzian signature and there are roots $\alpha$ of the algebra with norms $\alpha^2=2-2k$ for $k\in \mathbb{N}_0$. The roots with $\alpha^2=2$ are called real roots and they have multiplicity one; all others are imaginary and have higher multiplicity.

The subalgebra $\ke$ is generated by the `compact' combinations
\begin{align}\label{xi}
x_i =e_i - f_i
\end{align}
which are invariant under the Cartan-Chevalley involution
\begin{equation}
\omega(e_i) = - f_i \; , \quad
\omega(f_i) = - e_i \;,\quad
\omega(h_i) = - h_i
\end{equation}
The relations satisfied by these elements are in general not homogeneous (unlike the 
standard relations in the Chevalley--Serre presentation for the $e_i$ and $f_i$ above). Depending on whether two nodes $i$ and $j$ are connected by a line in the Dynkin diagram or not one has two cases
\begin{align}
\label{BRels}
\left[x_i, x_j \right] &=0 &&\textrm{if $i$ and $j$ are not connected}\nn\\
\left[x_i, \left[ x_i,x_j \right]\right] + x_j &=0 &&\textrm{if $i$ and $j$ are connected}
\end{align}
We will refer to these as the Berman--Serre relations; these relations were studied in a more general context in Ref.~\cite{Berman}. The algebra $\ke$ is then defined as the free Lie algebra over the
generators $\{x_i\}$  subject to the relations \eqref{BRels}. The task of finding representations of $\ke$ is  tantamount to finding matrices or operators that satisfy these relations. 

Since all simple generators $x_i$ are associated with real simple roots (of multiplicity one) one can also rephrase these relations more generally for any real roots by considering a generator $J(\alpha)$ for any (positive) real root $\alpha$. Using a basis of simple roots $\alpha_i$ of the root lattice one then has $x_i = J(\alpha_i)$ as particular case. The relations~\eqref{BRels} are then equivalent for real roots $\alpha$ and $\beta$ obeying $\alpha\cdot \beta\in \{-1,0,1\}$
\begin{align}
\label{BR2}
[J(\alpha) , J(\beta) ] &= \epsilon_{\alpha,\beta} J(\alpha+ \beta),&&\textrm{if $\alpha\cdot\beta=- 1$},&\nn\\
{}[J(\alpha) , J(\beta) ] &= -\epsilon_{\alpha,-\beta} J(\alpha- \beta),&&\textrm{if $\alpha\cdot\beta= +1$},&\nn\\
{}[J(\alpha) , J(\beta) ] &= 0,&&\textrm{if $\alpha\cdot\beta=0$},&
\end{align}
and $\epsilon_{\alpha,\beta}\in \{-1,1\}$ is a certain cocycle on the $\e$ root lattice that satisfies
\begin{align}
\epsilon_{\alpha,\beta} = - \epsilon_{\beta,\alpha} = - \epsilon_{-\alpha,-\beta},\quad \epsilon_{\alpha+\beta,-\beta}= \epsilon_{\alpha,\beta}.
\end{align}
The restriction on the inner product in the commutation is to make sure that $\alpha\mp \beta$ is a real root or no root at all, such that one does not have to worry about multiplicities from imaginary roots on the right-hand side. By contrast $\epsilon_{\alpha,\beta}$ can be defined for any pair of elements $(\alpha,\beta)$ of the root lattice.

To the root lattice of $\e$ one can also associate elements $\Gamma(\alpha)$ of the $\mf{so}(10)$ Clifford algebra of real $(32\times 32)$ matrices such that \cite{Kleinschmidt:2013eka}
\begin{align}
\label{GR2}
\Gamma(\alpha) \Gamma(\beta)  = \epsilon_{\alpha,\beta} \Gamma(\alpha+\beta) =-\epsilon_{\alpha,-\beta}\Gamma(\alpha-\beta).
\end{align}
With these rules it is then not hard to verify that 
\begin{align}
J(\alpha) = \frac12 \Gamma(\alpha)
\end{align}
provides a representation of $\ke$ for all real roots $\alpha$. This $32$-dimensional representation is known as the Dirac-spinor of $\ke$. By choosing a particular basis of the root lattice, called wall basis, one could exhibit\cite{Kleinschmidt:2013eka} that the $x_i$ for $i=1,\ldots,9$ are just the usual spin representation $x_i=\frac12 \Gamma^{i\,i+1}$ of $\mf{so}(10)$ but we will not use this here. 

\section{Tensors and spinors on Wheeler--DeWitt mini-superspace}

The space of diagonal spatial metrics in $11$ space-time dimensions is a Lorentzian 
ten-dimensional space in the Hamiltonian treatment of general relativity. 
This space is actually a finite-dimensional truncation of the full Wheeler--DeWitt 
`superspace' ({\it alias} the `moduli space of 10-geometries') 
to the finite-dimensional subspace of diagonal scale factors  (the negative direction
that renders this metric indefinite is associated with the scaling mode of the metric).
We choose a basis $e_\ta$ for 
 this ten-dimensional space ($\ta, \tb,\ldots=1,\ldots, 10$) with inner products
\begin{align}
e_\ta \cdot e_\tb = G_{\ta\tb}
\end{align}
where $G_{\ta\tb}$ is the Lorentzian DeWitt superspace metric restricted to the space of
metric scale factors; more explicitly, it follows from the Einstein--Hilbert action that
\begin{equation}\label{Gab}
G_{\ta\tb} = \delta_{\ta\tb} - 1 \quad \Rightarrow \qquad
G^{\ta\tb} = \delta_{\ta\tb} - \frac19
\end{equation}
This Lorentzian space can be identified with the Lorentzian space spanned by the roots of $\e$. 
In the remainder we do not require the explicit form of $G_{\ta\tb}$ of \eqref{Gab}.

Our ansatz for fermionic representations of $\ke$ then consists in considering tensor-spinors $\phi^{\ta_1\ldots \ta_n}_A = \phi^{(\ta_1\ldots \ta_n)}_A$ that are completely symmetric in their $n$ tensor indices and also carry a spinor index $A=1,\ldots, 32$ of $\mf{so}(10)$. The Dirac-spinor discussed in the preceding section then simply corresponds to $n=0$. We will also consider the case when $\phi^{\ta_1\ldots\ta_n}_A$ is traceless in its tensor indices. Since the tensor indices are those of a Lorentzian $\mf{so}(1,9)$ space while the spinor index belongs to the Euclidean $\mf{so}(10)$ subalgebra of $\ke$ our approach could be termed hybrid. Certainly one cannot take simple $\Gamma$-traces of $\phi^{\ta_1\ldots\ta_n}_A$ because $\ta,\tb,\dots$ are not 
SO(10) indices, so the only option to render the tensor-spinor irreducible is to make it
traceless in its indices $\ta_1,\ta_2,\dots$.

The generators $J(\alpha)$ of $\ke$ are then given by combinations of an object acting on the tensor indices and gamma matrices acting on the spinor index. More precisely, we make the ansatz
\begin{align}
J(\alpha)\phi^{\ta_1\ldots \ta_n}_A = -2 X(\alpha)^{\ta_1\ldots \ta_n}{}_{\tb_1\ldots\tb_n} \Gamma(\alpha)_{AB} \phi^{\tb_1\ldots \tb_n}_B.
\end{align}
Due to the known properties~\eqref{BR2} of the $\Gamma(\alpha)$ under commutation, checking the consistency relations~\eqref{GR2} then can be reduced to checking the following conditions on the tensors $X(\alpha)$ for real roots \cite{Kleinschmidt:2013eka}
\begin{align}
\label{BR3}
\left\{ X(\alpha)\,,\, X(\beta) \right\} &= \frac12  X(\alpha\pm \beta),&&\textrm{if $\alpha\cdot\beta=\mp 1$},&\nn\\
\left[ X(\alpha)\,,\, X(\beta) \right] &= 0,&&\textrm{if $\alpha\cdot\beta=0$}.&
\end{align}
Note that there is no $\epsilon_{\alpha,\beta}$ in these relations as it is already taken care of by the $\Gamma(\alpha)$. The Dirac-spinor corresponds to the solution $X(\alpha) = \frac14$ to these equations.

Another $\ke$ representation that has been known from supergravity considerations is the case $n=1$ that corresponds to the $D=11$ gravitino and has dimension $320$~\cite{Damour:2005zs,deBuyl:2005sch,Damour:2006xu}. In our language it corresponds to the solution
\begin{align}
X(\alpha)^\ta{}_\tb = -\frac12 \alpha^\ta\alpha_\tb + \frac14 \delta^\ta_\tb,
\end{align}
where $\alpha^\ta$ are the components of the root $\alpha$ with respect to the basis $e_\ta$, i.e., $\alpha=\sum_{\ta} \alpha^\ta e_\ta$. `Typewriter font' indices are raised and lowered with the Lorentzian $G_{\ta\tb}$.

For the gravitino (or vector-spinor) one can find a rewriting in terms of pure $\mf{so}(10)$ representation by letting\cite{Damour:2009zc}
\begin{align}
\psi^a_A = \sum_B \Gamma^a_{AB} \phi^a_B\quad\quad\textrm{(no sum on $a$)}.
\end{align}
The object on the left is then a standard vector-spinor of $\mf{so}(10)$. A similar simple and explicit rewriting into $\mf{so}(10)$ representations is not known for the new representations we discuss below.

We also note that due to the unfaithfulness of the representations, one obtains (infinite-dimensional) ideals in $\ke$, leading to the result that $\ke$ is not a simple Lie algebra. The quotient Lie algebras $\mf{q}$ of $\ke$ by the ideals of the ${\bf 32}$ and ${\bf 320}$ have been analysed and are given by $\mf{q}_{\bf 32} \cong \mf{so}(32)$ and $\mf{q}_{\bf 320}\cong \mf{so}(288,32)$. It may seem surprising that the `compact' $\ke$ admits a non-compact quotient in the ${\bf 320}$ representation but this is not a contradiction due to the infinite-dimensionality of $\ke$. For the higher spin representations below, the quotients have not been worked out.

\section{Higher spin representations}

In Ref.~\cite{Kleinschmidt:2013eka} two further solutions to~\eqref{BR3} were found that correspond to the values $n=2$ and $n=3$ (corresponding to spin $s=\frac52$ and
$s=\frac72$, respectively\cite{fn:spin}).
These representations go beyond supergravity as there appears 
to be no supergravity model from which they would be derivable. For spin $s=\frac52$
($n=2$) the corresponding tensors are given by
\begin{align}
\label{X52}
X(\alpha)^{\ta_1\ta_2}{}_{\tb_1\tb_2} = \frac12 \alpha^{\ta_1} \alpha^{\ta_2} \alpha_{\tb_1}\alpha_{\tb_2} -\alpha^{(\ta_1}_{\phantom{\ta_1)}} \delta^{\ta_2)}_{(\tb_1} \alpha_{\tb_2)}^{\phantom{\ta_1)}} + \frac14 \delta^{(\ta_1}_{\tb_1} \delta^{\ta_2)}_{\tb_2}
\end{align}
and for $n=3$ (spin-$\frac72$) by
\begin{align}
\label{X72}
X(\alpha)^{\ta_1\ta_2\ta_3}{}_{\tb_1\tb_2\tb_3}& =-\frac13 \alpha^{\ta_1} \alpha^{\ta_2}\alpha^{\ta_3} \alpha_{\tb_1}\alpha_{\tb_2}\alpha_{\tb_3}
+\frac32 \alpha^{(\ta_1} \alpha^{\ta_2}\delta^{\ta_3)}_{(\tb_1} \alpha_{\tb_2}\alpha_{\tb_3)}
-\frac32\alpha^{(\ta_1}\delta^{\ta_2}_{(\tb_1} \delta^{\ta_3)}_{\tb_2\phantom{)}}\alpha_{\tb_3)}^{\phantom{)}}&\nn\\
&\quad +\frac14 \delta^{(\ta_1}_{(\tb_1}\delta^{\ta_2\phantom{)}}_{\tb_2\phantom{)}}\delta^{\ta_3)}_{\tb_3)}
+\frac1{12}(2-\sqrt{3}) \alpha^{(\ta_1}  G^{\ta_2\ta_3)}G_{(\tb_1\tb_2}\alpha_{\tb_3)}&\\
&\quad+\frac1{12}(-1+\sqrt{3})\left( \alpha^{(\ta_1}\alpha^{\ta_2}\alpha^{\ta_3)}G_{(\tb_1\tb_2}\alpha_{\tb_3)}
+\alpha^{(\ta_1}  G^{\ta_2\ta_3)}\alpha_{(\tb_1}\alpha_{\tb_2}\alpha_{\tb_3)}\right).\nn
\end{align}
These expressions can be found and verified analytically. We have also extended the search for solutions of this type for $n\leq 10$ with the ansatz above but have not found any additional 
solutions so far.

The spin-$\frac52$ solution as given is of dimension $\frac{10\times 11}{2} \times 32= 1760$. It turns out that this representation is reducible since the subspace spanned by the trace $G_{\ta\tb} \phi^{\ta\tb}_A$ is invariant. This trace transforms in the spin-$\frac12$ representation of dimension $32$, leaving an irreducible $1728$-dimensional representation of $\ke$.
By contrast, the spin-$\frac72$ representation of dimension $\frac{10\times 11\times 12}{6}\times 32 = 7040$ is irreducible as given.

In the next two sections, we investigate further properties of the new higher spin representations. 

\section{Projectors and Weyl group action}

The $\ke$ generators $J(\alpha)$ are defined for all positive roots $\alpha$ of $\e$. As the roots $\alpha$ are space-like elements in a Lorentzian ten-dimensional space, they have a stabiliser of type $\mf{so}(1,8)\subset \mf{so}(1,9)$. This stability algebra can be used to decompose the `polarisation tensor' $X(\alpha)$ into irreducible pieces for a fixed $\alpha$. The irreducible 
$\mf{so}(1,8)$ terms are given by projectors $\Pi^{(j)}(\alpha)$, 
 such that tensor $X(\alpha)$ can be expressed in terms of these projectors.\cite{fn:proj} This rewriting 
greatly facilitates the exponentiation of the corresponding matrices, and
will make it easy to work out the exponentiated (Weyl) group actions.

\subsection{Projectors for spin-$\frac52$}

For $n=2$, the polarisation tensor $X(\alpha)^{\ta_1\ta_2}{}_{\tb_1\tb_2}$ lies in the reducible ${\bf 55}$ of $\mf{so}(1,9)$, where we work for simplicity with the reducible representation of dimension $1760$ given in~\eqref{X52}. The decomposition of $X(\alpha)$ under the regularly embedded $\mf{so}(1,8)$ is
\begin{align}
{\bf 55} \rightarrow {\bf 54} \oplus {\bf 1} \rightarrow ({\bf 44} \oplus {\bf 9} \oplus {\bf 1'}) \oplus {\bf 1}.
\end{align}
The splitting of the singlets here has been done in such a way that ${\bf 1}$ corresponds to the $\mf{so}(1,9)$ singlet corresponding to the trace with $G_{\ta\tb}$.
One can check that the following are complete orthonormal projectors on the various pieces
\begin{align}
\Pi^{({\bf 44})}(\alpha)^{\ta_1\ta_2}{}_{\tb_1\tb_2} &= \frac29 \alpha^{\ta_1} \alpha^{\ta_2} \alpha_{\tb_1}\alpha_{\tb_2}
   - \alpha^{(\ta_1} \delta^{\ta_2)}_{(\tb_1} \alpha_{\tb_2)}
   + \delta^{(\ta_1}_{\tb_1} \delta^{\ta_2)}_{\tb_2} &\nonumber\\
&\quad+\frac1{18} \left(\alpha^{\ta_1} \alpha^{\ta_2} G_{\tb_1\tb_2} + G^{\ta_1\ta_2} \alpha_{\tb_1} \alpha_{\tb_2}\right)
   -\frac19 G^{\ta_1\ta_2} G_{\tb_1\tb_2},&\nn\\
\Pi^{({\bf 9})}(\alpha)^{\ta_1\ta_2}{}_{\tb_1\tb_2} &= -\frac12 \alpha^{\ta_1} \alpha^{\ta_2} \alpha_{\tb_1}\alpha_{\tb_2} +  \alpha^{(\ta_1} \delta^{\ta_2)}_{(\tb_1} \alpha_{\tb_2)},\\
\tilde{\Pi}^{({\bf 1})}(\alpha)^{\ta_1\ta_2}{}_{\tb_1\tb_2} &= \frac1{10}G^{\ta_1\ta_2} G_{\tb_1\tb_2},&\nn\\
\tilde{\Pi}^{({\bf 1'})}(\alpha)^{\ta_1\ta_2}{}_{\tb_1\tb_2} &= \frac{5}{18} \alpha^{\ta_1} \alpha^{\ta_2} \alpha_{\tb_1}\alpha_{\tb_2} -\frac1{18} \left(\alpha^{\ta_1} \alpha^{\ta_2} G_{\tb_1\tb_2} + G^{\ta_1\ta_2} \alpha_{\tb_1} \alpha_{\tb_2}\right)&\nn\\
&\quad    +\frac1{90} G^{\ta_1\ta_2} G_{\tb_1\tb_2}.\nn
\end{align}
In terms of these, the tensor $X(\alpha)$ takes the form
\begin{align}
X(\alpha)^{\ta_1\ta_2}{}_{\tb_1\tb_2} 
&= \left( \frac14 \tilde{\Pi}^{({\bf 1})}(\alpha) +\frac14\tilde{\Pi}^{({\bf 1'})}(\alpha)-\frac34 \Pi^{({\bf 9})}(\alpha) +\frac14 \Pi^{({\bf 44})}(\alpha) \right){}^{\ta_1\ta_2}{}_{\tb_1\tb_2}. 
\end{align}
What is important here is that the coefficients of all projectors are of the form $\frac{2k+1}{4}$ with $k\in\mathbb{Z}$. This implies that when one constructs the `Weyl group' generator
\begin{align}
w_\alpha = e^{\frac{\pi}{2} J(\alpha)}
\end{align}
acting in the representation is idempotent in the eighth power. Weyl group has been put into inverted commas above because this is more correctly an element of a covering of the Weyl group that has been dubbed the spin-extended Weyl group~\cite{Damour:2009zc,GHKW}. Acting on spinor representations, the characteristic feature is that only the eighth power $w_\alpha^8=1\!\!1$ whereas one normally has the fourth power for the covering of the Weyl on bosonic representations~\cite{Kac}.

\subsection{Projectors for spin-$\frac72$}

In this case, the polarisation tensor $X(\alpha)^{\ta_1\ta_2\ta_3}{}_{\tb_1\tb_2\tb_3}$ is in the (reducible) totally symmetric ${\bf 220}$ of $\mf{so}(1,9)$. This decomposes under $\mf{so}(1,8)$ as
\begin{align}
{\bf 220} \rightarrow {\bf 210} \oplus {\bf 10} \rightarrow \left( {\bf 156} \oplus {\bf 44} \oplus {\bf 9} \oplus {\bf 1}\right) \oplus \left({\bf 9} \oplus  {\bf 1}\right).
\end{align}
There are two singlets and two vectors of $\mf{so}(1,8)$ appearing in the decomposition and some associated freedom in constructing the orthonormal projectors. We choose a particular combination of these representations as follows~\cite{ProjNote}
\begin{align}
\Pi^{({\bf 156})}(\alpha)^{\ta_1\ta_2\ta_3}{}_{\tb_1\tb_2\tb_3} &= -\frac1{11} \alpha^{\ta_1} \alpha^{\ta_2}\alpha^{\ta_3} \alpha_{\tb_1}\alpha_{\tb_2}\alpha_{\tb_3}
  +\frac{15}{22}\alpha^{(\ta_1} \alpha^{\ta_2} \delta^{\ta_3)}_{(\tb_1} \alpha_{\tb_2}\alpha_{\tb_3)}\nn\\
&\quad   -\frac32\alpha^{(\ta_1} \delta^{\ta_2}_{(\tb_1} \delta^{\ta_3)}_{\tb_2} \alpha_{\tb_3)}
 +\delta^{(\ta_1}_{(\tb_1} \delta^{\ta_2}_{\tb_2} \delta^{\ta_3)}_{\tb_3)}\nn\\
&\quad   -\frac3{44}\left(\alpha^{\ta_1} \alpha^{\ta_2} \alpha^{\ta_3} \alpha_{(\tb_1} G_{\tb_2\tb_3)} + \alpha^{(\ta_1} G^{\ta_2\ta_3)} \alpha_{\tb_1}\alpha_{\tb_2}\alpha_{\tb_3}\right)\nn\\
&\quad + \frac3{22}\left( \alpha^{(\ta_1} \alpha^{\ta_2} \delta^{\ta_3)}_{(\tb_1} G_{\tb_2\tb_3)} 
  + G^{(\ta_1\ta_2} \delta^{\ta_3)}_{(\tb_1} \alpha_{\tb_2} \alpha_{\tb_3)}\right) \nn\\
&\quad  -\frac3{11} G^{(\ta_1\ta_2} \delta^{\ta_3)}_{(\tb_1} G_{\tb_2\tb_3)} +\frac3{22}\alpha^{(\ta_1} G^{\ta_2\ta_3)} G_{(\tb_1\tb_2} \alpha_{\tb_3)}\nn\\
\Pi^{({\bf 44})}(\alpha)^{\ta_1\ta_2\ta_3}{}^{\tb_1\tb_2\tb_3} &= \frac13 \alpha^{\ta_1} \alpha^{\ta_2}\alpha^{\ta_2} \alpha_{\tb_1}\alpha_{\tb_2}\alpha_{\tb_3}
  -\frac32\alpha^{(\ta_1} \alpha^{\ta_2} \delta^{\ta_3)}_{(\tb_1} \alpha_{\tb_2}\alpha_{\tb_3)}\nn\\
&\quad  +\frac32\alpha^{(\ta_1} \delta^{\ta_2}_{(\tb_1} \delta^{\ta_3)}_{\tb_2} \alpha_{\tb_3)}
   -\frac1{6}\alpha^{(\ta_1} G^{\ta_2\ta_3)} G_{(\tb_1\tb_2} \alpha_{\tb_3)}\nn\\
&\quad 
   +\frac1{12}\left(\alpha^{\ta_1} \alpha^{\ta_2} \alpha^{\ta_3} \alpha_{(\tb_1} G_{\tb_2\tb_3)} + \alpha^{(\ta_1} G^{\ta_2\ta_3)} \alpha_{\tb_1}\alpha_{\tb_2}\alpha_{\tb_3}\right)
\end{align}
for the (unique) two biggest representations,
\begin{align}
\Pi^{({\bf 9})}(\alpha)^{\ta_1\ta_2\ta_3}{}_{\tb_1\tb_2\tb_3} &= -\frac9{22}\alpha^{\ta_1} \alpha^{\ta_2}\alpha^{\ta_3} \alpha_{\tb_1}\alpha_{\tb_2}\alpha_{\tb_3}
  +\frac{9}{11}\alpha^{(\ta_1} \alpha^{\ta_2} \delta^{\ta_3)}_{(\tb_1} \alpha_{\tb_2}\alpha_{\tb_3)}\nn\\
&\quad
   +\frac3{44}\left(\alpha^{\ta_1} \alpha^{\ta_2} \alpha^{\ta_3} \alpha_{(\tb_1} G_{\tb_2\tb_3)} + \alpha^{(\ta_1} G^{\ta_2\ta_3)} \alpha_{\tb_1}\alpha_{\tb_2}\alpha_{\tb_3}\right)\nn\\
&\quad - \frac3{22}\left( \alpha^{(\ta_1} \alpha^{\ta_2} \delta^{\ta_3)}_{(\tb_1} G_{\tb_2\tb_3)}  + G^{(\ta_1\ta_2} \delta^{\ta_3)}_{(\tb_1} \alpha_{\tb_2} \alpha_{\tb_3)}\right) \nn\\
&\quad  +\frac1{44} G^{(\ta_1\ta_2} \delta^{\ta_3)}_{(\tb_1} G_{\tb_2\tb_3)} 
  -\frac1{88}\alpha^{(\ta_1} G^{\ta_2\ta_3)} G_{(\tb_1\tb_2} \alpha_{\tb_3)}\nn\\
\Pi^{({\bf 9}')}(\alpha)^{\ta_1\ta_2\ta_3}{}_{\tb_1\tb_2\tb_3} &=   \frac1{4} G^{(\ta_1\ta_2} \delta^{\ta_3)}_{(\tb_1} G_{\tb_2\tb_3)} 
  -\frac1{8}\alpha^{(\ta_1} G^{\ta_2\ta_3)} G_{(\tb_1\tb_2} \alpha_{\tb_3)}
\end{align}
for the vectors and finally for the singlets
\begin{align}
\Pi^{({\bf 1})}(\alpha)^{\ta_1\ta_2\ta_3}{}_{\tb_1\tb_2\tb_3} &= \frac1{12} \alpha^{\ta_1} \alpha^{\ta_2}\alpha^{\ta_3} \alpha_{\tb_1}\alpha_{\tb_2}\alpha_{\tb_3}&\nn\\
&\quad   +\frac1{24}(-1-\sqrt{3})\left(\alpha^{\ta_1} \alpha^{\ta_2} \alpha^{\ta_3} \alpha_{(\tb_1} G_{\tb_2\tb_3)} + \alpha^{(\ta_1} G^{\ta_2\ta_3)} \alpha_{\tb_1}\alpha_{\tb_2}\alpha_{\tb_3}\right)&\nn\\
&\quad   +\frac1{24}(2+\sqrt{3})\alpha^{(\ta_1} G^{\ta_2\ta_3)} G_{(\tb_1\tb_2} \alpha_{\tb_3)}\nn\\
\Pi^{({\bf 1}')}(\alpha)^{\ta_1\ta_2\ta_3}{}_{\tb_1\tb_2\tb_3} &= \frac1{12} \alpha^{\ta_1}\alpha^{\ta_2}\alpha^{\ta_3} \alpha_{\tb_1}\alpha_{\tb_2}\alpha_{\tb_3}&\nn\\
&\quad   +\frac1{24}(-1+\sqrt{3})\left(\alpha^{\ta_1} \alpha^{\ta_2} \alpha^{\ta_3} \alpha_{(\tb_1} G_{\tb_2\tb_3)} + \alpha^{(\ta_1} G^{\ta_2\ta_3)} \alpha_{\tb_1}\alpha_{\tb_2}\alpha_{\tb_3}\right)&\nn\\
&\quad   +\frac1{24}(2-\sqrt{3})\alpha^{(\ta_1} G^{\ta_2\ta_3)} G_{(\tb_1\tb_2} \alpha_{\tb_3)}
\end{align}
The tensor $X(\alpha)$ reads as follows in this basis
\begin{align}
X(\alpha)^{\ta_1\ta_2\ta_3}{}_{\tb_1\tb_2\tb_3} 
&= \Bigg( \frac54\Pi^{({\bf 1})}(\alpha)-\frac34\Pi^{({\bf 1}')}(\alpha) +\frac14\Pi^{({\bf 9})}(\alpha)+\frac14 \Pi^{({\bf 9}')}(\alpha) &\nn\\
&\quad\quad - \frac34\Pi^{({\bf 44})}(\alpha)+\frac14 \Pi^{({\bf 156})}(\alpha) \Bigg){}^{\ta_1\ta_2\ta_3}{}_{\tb_1\tb_2\tb_3}.
\end{align}
Again, it is important that the coefficients of all the orthonormal projectors are of the form $\frac{2k+1}{4}$ such that we are dealing with a genuine fermionic representation of $\KE$. 

\section{Branching under subalgebras}

The infinite-dimensional Lie algebra $\ke$ has infinitely many finite-dimensional 
subalgebras.\cite{fn:sub}
Of these are of particular interest to us the following, all of which can be obtained by deleting a single node from the $\e$ Dynkin diagram:\\[2mm]
\begin{tabular}{p{3mm}p{30mm}p{30mm}p{50mm}}
$(a)$ & $\mf{so}(10)$ & deleting node $10$ & SUGRA in $D=11$\\
$(b)$ & $\mf{so}(2)\oplus \mf{so}(16)$ & deleting node $2$ & SUGRA in $D=3$\\
$(c)$ & $\mf{so}(9)\oplus \mf{so}(2)$ & deleting node $8$ & IIB SUGRA in $D=10$\\
$(d)$ & $\mf{so}(9)\oplus \mf{so}(9)$ & deleting node $9$ & Doubled SUGRA in $D=10$
\end{tabular}
\vspace{2mm}

The last case requires some explanation. In Ref.~\cite{Kleinschmidt:2004dy} the decomposition of $\e$ under its $\mf{so}(9,9)$ subalgebra was studied and shown to correspond to both type IIA and type IIB theory since the Ramond--Ramond potentials occurred in a spinor representation of  $\mf{so}(9,9)$ that can be read either as all even or all odd forms; similarly, the fermions arrange themselves correctly for the two theories~\cite{Kleinschmidt:2006tm}. In investigations of double field theory the same structure appears~\cite{Jeon:2011vx} and we have therefore dubbed this T-duality agnostic decomposition as `doubled SUGRA.'

There are some additional subtleties associated with the global assignment of fermionic and bosonic representations at the group level. More precisely, the $\mf{so}(16)$ is the Lie algebra of $Spin(16)/\mathbb{Z}_2$. The $\mathbb{Z}_2$ is not diagonally embedded in the center $\mathbb{Z}_2\times \mathbb{Z}_2$ but as one of the factors; this entails that the representations ${\bf 16}_v$ and 
${\bf 128}_c$ are spinorial (that is, they transform with a factor $(-1)$ upon rotation by $2\pi$),
whereas the ${\bf 128}_s$ is tensorial~\cite{Keurentjes:2003yu}. Moreover, the ${\bf 16}$ spinor of $Spin(9)=[Spin(9)\times Spin(9)]_{\textrm{diag}}$ is identified with the (spinorial) ${\bf 16}_v$ of $Spin(16)$. The diagonal $Spin(9)$ also lies as a regular subgroup in $Spin(9)$ as it corresponds to the dimensional reduction from $D=11$ to $D=10$ (over a spatial direction).

The decompositions of the spin-$\frac12$ and spin-$\frac32$ representations were already given in Ref.~\cite{Kleinschmidt:2006tm}, while the decompositions of the new higher spin 
representations under the various subalgebras have not been given previously.
To find the relevant decompositions for spin-$\frac52$ and spin-$\frac72$ 
is actually rather involved, and can only be done on a computer. The
main problem here is that the $\KE$ representations are not highest or lowest weight representations (it is not even clear whether $\KE$ admits any analog of such representations),
so the customary tools of representation theory cannot be applied. However, the subrepresentations obtained after descending to any finite-dimensional subgroup 
{\em are} highest or lowest
weight representations, so given any of the above subgroups, one must first identify the 
corresponding highest or lowest weights. For instance, for the spin-$\frac72$ representation 
this requires (amongst other things) the (simultaneous) diagonalisation of various 7040 $\times$ 7040 matrices. 
It seems clear that for yet higher dimensional realisations such a procedure would become impractical very quickly unless better methods are developed.

\subsection{Branching the spin $s=\frac12$ and $s=\frac32$ representations}

These were already understood in previous work \cite{Kleinschmidt:2006tm}.
The fractions $\frac12$ and $\frac32$ in the decompositions $(b)$ and $(c)$ below correspond to the $\mf{so}(2)\cong\mf{u}(1)$ charges. In these cases all the representations form doublets of $\mf{so}(2)$ that can also be thought of as complex one-dimensional representations of $\mf{u}(1)$. This has to be taken into account when checking the dimension count of the decompositions. 
\begin{align}
{\bf 32}\quad &\stackrel{a}\longrightarrow \quad {\bf 32}   \nn\\[2mm]
        &\stackrel{b}{\longrightarrow} \quad \left(\frac12, {\bf 16}_v\right)  \nn\\[2mm]
        &\stackrel{c}{\longrightarrow} \quad \left({\bf 16},\frac12\right)\nn\\[2mm]
        &\stackrel{d}{\longrightarrow} \quad ({\bf 16},{\bf 1}) \oplus
              ({\bf 1},{\bf 16}) 
 \end{align}
and\allowdisplaybreaks{
\begin{align}
{\bf 320} \quad &\stackrel{a}\longrightarrow \quad {\bf 288} \oplus {\bf 32} \nn\\[2mm]
     &\stackrel{b}\longrightarrow \quad \left( \frac12,{\bf 128}_c\right)
\oplus \left(\frac12, {\bf16}_v\right)  \oplus \left(\frac32, {\bf16}_v\right)   \nn\\[2mm]
        &\stackrel{c}{\longrightarrow} \quad \left({\bf 16}, \frac32\right) \oplus
              \left({\bf 128}, \frac12\right) \oplus 
              \left({\bf 16}, \frac12\right) 
\nn\\[2mm]
  &\stackrel{d}{\longrightarrow} \quad \left({\bf 9}, {\bf 16}\right)\oplus \left({\bf 16}, {\bf 9} \right)
      \oplus  \left({\bf 1}, {\bf 16} \right) \oplus \left({\bf 16}, {\bf 1}\right)     
 \end{align}
Since these are the `physical' fermions of maximal supergravity, let us briefly comment on their interpretation.}

The ${\bf 32}$ representation of $K(\mf{e}_{10})$ corresponds to the $32$ supersymmetry generators of maximal supergravity. We see that in the decomposition $(a)$ relevant for $D=11$ supergravity one obtains a single generator consistent with $\mathcal{N}=1$ supersymmetry. In the decomposition $(b)$ one obtains an $\mf{so}(2)$ doublet of sixteen generators (in the vector of $\mf{so}(16)$; the $\mf{so}(2)$ corresponds to the spatial part of the $\mf{so}(1,2)$ Lorentz symmetry of which the doublet is the irreducible spinor and the sixteen components correspond to  $\mf{so}(16)$ R-symmetry of maximal $\mathcal{N}=16$ supersymmetry in $D=3$ dimensions. The decomposition $(c)$ gives an $\mf{so}(2)$ R-symmetry doublet of spinors of the spatial $\mf{so}(9)$ Lorentz symmetry in $D=10$ in agreement with the supersymmetry generators of chiral type IIB supergravity. More specifically,
the appearance of this U(1) effectively `complexifies' the SO(9) representation, in line
with the chirality of the type IIB fermions. The last decomposition $(d)$ is consistent 
with a type IIA formulation of doubled supergravity~\cite{Kleinschmidt:2004dy,Jeon:2011vx}.

The decompositions of the ${\bf 320}$ representation of $K(\mf{e}_{10})$ can be interpreted similarly~\cite{Kleinschmidt:2006tm}. For example, the decomposition $(b)$ gives the $128$ physical fermions in $D=3$ together with components associated with the non-propagating gravitino that is needed when formulating $\mathcal{N}=16$ supergravity in $D=3$. We also note again that in the type IIB decomposition $(c)$ one always obtains doublets of the R-symmetry $\mf{so}(2)$, in accord
with the chirality of the underlying fermionic multiplets.

\subsection{Branching of the spin-$\frac52$ representation}

The decomposition under the various subalgebras is
\begin{align}
{\bf 1728} &\stackrel{a}{\longrightarrow} {\bf 1120} \,\oplus \, 2\times {\bf 288} 
      \,\oplus \, {\bf32}  \nn\\[2mm]
        &\stackrel{b}{\longrightarrow} \left(\frac12 , {\bf 560}_v \right) \oplus 
        \left(\frac12 ,{\bf 128}_c \right) 
              \oplus 2\times \left(\frac12 , {\bf 16}_v \right) \oplus \left(\frac32 , {\bf 128}_c, \right) 
              \oplus \left(\frac32 , {\bf 16}_v \right)  \nn\\[2mm]
        &\stackrel{c}{\longrightarrow} \left({\bf 432},\frac12\right) \oplus 
        2\times \left({\bf 128},\frac12\right) \oplus 2\times \left({\bf 16},\frac12\right) 
              \oplus \left({\bf 128},\frac32\right) \oplus \left({\bf 16},\frac32\right)\nn\\[2mm]
        &\stackrel{d}{\longrightarrow} ({\bf 36}, {\bf 16})\,\oplus\,
           ({\bf 16}, {\bf 36})\,\oplus\,
           ({\bf 9}, {\bf 16})\,\oplus\,
           ({\bf 16},{\bf 9})\,\oplus\nn\\[2mm]
&\quad\quad\oplus           ({\bf 128}, {\bf 1})\,\oplus\,
           ({\bf 1}, {\bf 128})\,\oplus\,
           ({\bf 1} , {\bf 16}) \, \oplus\,
              ({\bf 16}, {\bf 1}) 
 \end{align}
From the $\mf{so}(10)$ decomposition we see that the space-time spin of this $\ke$ representation is not really higher than $3/2$ since the ${\bf 1120}$ corresponds to an \textit{anti-}symmetric tensor-spinor of $\mf{so}(10)$ with two tensor indices. The ${\bf 560}_v$ of $\mf{so}(16)$ that arises is the anti-symmetric three-form. Similar to the ${\bf 16}_v$ discussed above, this is actually a spinorial representation with the correct assignment when lifted to the group $Spin(16)/\mathbb{Z}_2$. The ${\bf 432}$ of $\mf{so}(9)$ that arises in case $(c)$ is the tensor-spinor with two antisymmetric indices.

\subsection{Branching of the spin-$\frac72$ representation}

Under the subalgebras listed above, the $\ke$ spin-$\frac72$ representation 
of dimension $7040$ decomposes as
\begin{align}
{\bf 7040} &\stackrel{a}{\longrightarrow} {\bf 2400} \oplus {\bf 1440} \oplus 2 \times {\bf 1120} \oplus 3\times {\bf 288}  \oplus 3\times {\bf 32} \nn\\[2mm]
              &\stackrel{b}{\longrightarrow} \left(\frac12,{\bf 1920}_s\right) 
                   \oplus \left(\frac32,{\bf 560}_v\right) \oplus \left(\frac12,{\bf 560}_v\right)
                   \oplus \left(\frac32,{\bf 128}_c\right) \oplus 2\times \left(\frac12,{\bf 128}_c\right)\nn\\
                 &\quad\quad  \oplus \left(\frac52,{\bf 16}_v\right) \oplus 2\times \left(\frac32,{\bf 16}_v\right) 
                     \oplus 3\times \left(\frac12,{\bf 16}_v\right)
               \nn\\[2mm]
              &\stackrel{c}{\longrightarrow}
                  \left({\bf 768},\frac12\right) \oplus \left({\bf 576},\frac12\right) 
                  \oplus \left({\bf 432},\frac32\right) \oplus 2\times \left({\bf 432},\frac12\right)\nn\\
               &\quad\quad     \oplus 2\times \left({\bf 128},\frac32\right) \oplus 4\times \left({\bf 128},\frac12\right)
                  \oplus \left({\bf 16},\frac52\right) \oplus 2\times \left({\bf 16},\frac32\right)
                  \oplus 4\times \left({\bf 16},\frac12\right)\nn\\[2mm]
              &\stackrel{d}{\longrightarrow}  ({\bf 128}, {\bf 9}) \oplus ({\bf 128}, {\bf 1})
                  \oplus ({\bf 16}, {\bf 84}) \oplus ({\bf 16}, {\bf 36}) 
                  \oplus 2\times ({\bf 16}, {\bf 9}) \oplus 2\times ({\bf 16}, {\bf 1})\nn\\[2mm]
                &\quad\quad\oplus ({\bf 9}, {\bf 128}) \oplus ({\bf 1}, {\bf 128})
                  \oplus ({\bf 84}, {\bf 16}) \oplus ({\bf 36}, {\bf 16}) 
                  \oplus 2\times ({\bf 9}, {\bf 16}) \oplus 2\times ({\bf 1}, {\bf 16})
 \end{align}
As already mentioned, it is a non-trivial task to work out these decompositions in practice.
As a further test we have also checked that the
further decompositions of the $\mf{so}(2) \oplus \mf{so}(16)$ and $\mf{so}(10)$ representations under
their common $\mf{so}(8)$ subalgebra coincide (for $\mf{so}(16)$ this subalgebra is obtained after
descending first to the diagonal subalgebra $[\mf{so}(8) \oplus \mf{so}(8)]_{\rm diag}$). Similarly, there is another $\mf{so}(8)$ that is common to the $\mf{so}(10)$ decomposition $(a)$, to the type IIB decomposition $(c)$ and to the $\mf{so}(9)\oplus \mf{so}(9)$ decomposition in $(d)$ and that corresponds to the spatial rotations of maximal $D=9$ supergravity. The further branching of $(a)$, $(c)$ and $(d)$ to this common subgroup has been checked to be consistent. Moreover, we have verified that the common $\mf{so}(9)$ of the type IIB decomposition $(c)$ and the T-duality agnostic decomposition $(d)$ gives the same representations.
 
Let us finally highlight some new features arising here, that have no analog for
spin $s\leq \frac52$.
\begin{itemize}
\item  In the $\mf{so}(10)$ decomposition $(a)$ one sees the ${\bf 2400}$ that corresponds to a tensor-spinor that is antisymmetric in {\em three} tensor indices. The ${\bf 1440}$ is a tensor-spinor with two \textit{symmetric} tensor indices; since the $\mf{so}(10)$ is the spatial rotation group of $D=11$ supergravity, this means that the spin-$\tfrac72$ of $K(\mf{e}_{10})$ contains genuinely higher spin representations also from a space-time perspective! 
\item Under the $\mf{so}(2)\oplus \mf{so}(16)$ decomposition $(b)$ one finds the vector-spinor of $\mf{so}(16)$ with $1920$ components. Note that consistent with the spinorial nature of the $K(E_{10})$ representation it is the ${\bf 1920}_s$ where the spinorial double-valued aspect of $Spin(16)/\mathbb{Z}_2$ is carried by the vector index and not by the $s$-type spinor index.
\item The ${\bf 768}$ appearing in the $\mf{so}(9)\oplus \mf{so}(2)$ decomposition $(c)$ is the anti-symmetric three-form tensor-spinor of $\mf{so}(9)$. By contrast the ${\bf 576}$ is a tensor-spinor with two \textit{symmetric} tensor indices and therefore this $K(\mf{e}_{10})$ representation also contains fermionic higher spin fields from the type IIB perspective.
\item The ${\bf 84}$ in the $\mf{so}(9)\oplus\mf{so}(9)$ decomposition $(d)$ is the anti-symmetric three-form of $\mf{so}(9)$; the ${\bf 36}$ is the anti-symmetric two-form already encountered above. 
\end{itemize}
We also note that the $\mf{so}(2)$ eigenvalues can become larger and larger the bigger the $K(\mf{e}_{10})$ representation becomes. 

\section{Outlook}

There are two pressing questions arising out of our work. The first concerns the possible physical role of the new $\KE$ representations. In particular, one may wonder whether they are of relevance to overcoming the difficulties in constructing a supersymmetric $\E$ model that were encountered in~Ref.~\cite{Kleinschmidt:2014uwa}. It is conceivable that in order to make progress both 
the supersymmetry constraint and the propagating fermions will have to be assigned to representations of $\KE$ different from the ones used so far (and in particular incorporate
spatial gradients in one form or another). Let us also note that one can easily couple
the new fermion representations to the bosonic $\E/\KE$ sigma model, namely by adding
a Dirac-like term $\propto \Psi D_t \Psi$ to the bosonic action, where
$D_t \equiv \partial_t + \sum_{\alpha, r} Q^r(\alpha) J^r(\alpha)$ is the $\KE$ covariant
derivative, and $Q^r(\alpha)$ the $\ke$-connection as computed from the bosonic
sigma model in the standard way.
Of course, there remains the question whether one can define a new supersymmetry
that makes the combined action supersymmetric at least at low levels.

Secondly, the very existence of the two new higher spin representations for $s=\frac52$ and
$s=\frac72$ which cannot be explained from maximal supergravity, strongly suggests
that these constitute only the tip of the iceberg of the unexplored representation theory 
of $\ke$. Although our (limited) search for new examples has not been successful so far,
we expect there to exist an infinite tower of such realisations of higher and 
higher spin, which are less and less unfaithful with increasing spin, but which can occur
only at `sporadic' values of the spin, because the simultaneous decomposability 
under all the subgroups analysed in the foregoing section puts very tight constraints
on such new representations.\cite{fn:cons} We reiterate that working out these decompositions is currently a tedious task due to the lack of general methods for studying the representation theory of $\ke$.
An explicit construction of further examples and, more
ambitiously, a systematic understanding of their structure would afford an 
entirely new method to explore the root spaces associated with timelike imaginary
roots, and thus one of the main obstacles towards a better understanding of $\e$. One step forward might be the understanding of the decomposition of tensor products of $\ke$ representations.
We thus hope that our investigations help to clarify the structure of this enigmatic 
object and maybe also the elusive Kac--Moody algebra $\e$ itself. 

\section*{Acknowledgements}
AK would like to thank the organisers of the workshop on `Higher Spin Gauge Theories' for putting together a stimulating program as well as the Institute for Advanced Study of Nanyang Technical University for its generous and kind hospitality.

%\begin{appendix}[Optional Appendix Title]
%
%\section{Sample Appendix}
%Text...
%\end{appendix}

%\bibliographystyle{ws-rv-van}
%\bibliography{ws-rv-sample}

\begin{thebibliography}{99}        % for non BIBTeX users

\bibitem{Julia} B. Julia, ``Group Disintegrations,'' in: S. W. Hawking and M. Ro\v{c}ek (eds.), \textit{Superspace and Supergravity}, Proceedings of the Nuffield Work-shop, Cambridge, Eng., Jun 22--Jul 12, 1980, Cambridge University Press (Cambridge, 1981) 331--350; ``Kac--Moody Symmetry of Gravitation and Supergravity Theories,'' in: M. Flato, P. Sally and G. Zuckerman (eds.), \textit{Applications of Group Theory in Physics and Mathematical Physics} (Lectures in Applied Mathematics {\bf 21}), Am. Math. Soc. (Providence, 1985) 355--374, LPTENS 82/22.

\bibitem{Damour:2002cu}
  T.~Damour, M.~Henneaux and H.~Nicolai,
  ``E(10) and a 'small tension expansion' of M theory,''
  \doi{Phys.\ Rev.\ Lett.\  {\bf 89} (2002) 221601}{10.1103/PhysRevLett.89.221601}
  [\eprintth{hep-th/0207267}].
  %%CITATION = doi:10.1103/PhysRevLett.89.221601;%%

\bibitem{deBuyl:2005zy}
  S.~de Buyl, M.~Henneaux and L.~Paulot,
 ``Hidden symmetries and Dirac fermions,''
  \doi{Class.\ Quant.\ Grav.\  {\bf 22} (2005) 3595}{10.1088/0264-9381/22/17/018}
  [\eprintth{hep-th/0506009}].
  %%CITATION = doi:10.1088/0264-9381/22/17/018;%%

\bibitem{Damour:2005zs}
  T.~Damour, A.~Kleinschmidt and H.~Nicolai,
  ``Hidden symmetries and the fermionic sector of eleven-dimensional supergravity,''
  \doi{Phys.\ Lett.\ B {\bf 634} (2006) 319}{10.1016/j.physletb.2006.01.015}
  [\eprintth{hep-th/0512163}].
  %%CITATION = doi:10.1016/j.physletb.2006.01.015;%%

\bibitem{deBuyl:2005sch}
  S.~de Buyl, M.~Henneaux and L.~Paulot,
  ``Extended E(8) invariance of 11-dimensional supergravity,''
  \doi{JHEP {\bf 0602} (2006) 056}{10.1088/1126-6708/2006/02/056}
  [\eprintth{hep-th/0512292}].
  %%CITATION = doi:10.1088/1126-6708/2006/02/056;%%

\bibitem{Damour:2006xu}
  T.~Damour, A.~Kleinschmidt and H.~Nicolai,
  ``K(E(10)), Supergravity and Fermions,''
  \doi{JHEP {\bf 0608} (2006) 046}{10.1088/1126-6708/2006/08/046}
  [\eprintth{hep-th/0606105}].
  %%CITATION = doi:10.1088/1126-6708/2006/08/046;%%

\bibitem{Kleinschmidt:2007zd}
  A.~Kleinschmidt,
  ``Unifying R-symmetry in M-theory,''
  in: V. Sidoravi\v{c}ius (ed.) \textit{New Trends in Mathematical Physics}, 
  Proceedings of the XVth International Congress on Mathematical Physics, Springer (2009),
  [\eprintth{hep-th/0703262}].
  %%CITATION = HEP-TH/0703262;%%

\bibitem{Kleinschmidt:2004dy}
  A.~Kleinschmidt and H.~Nicolai,
  ``E(10) and SO(9,9) invariant supergravity,''
  \doi{JHEP {\bf 0407} (2004) 041}{10.1088/1126-6708/2004/07/041}
  [\eprintth{hep-th/0407101}].
  %%CITATION = doi:10.1088/1126-6708/2004/07/041;%%  

\bibitem{Kleinschmidt:2004rg}
  A.~Kleinschmidt and H.~Nicolai,
 ``IIB supergravity and E(10),''
  \doi{Phys.\ Lett.\ B {\bf 606} (2005) 391}{10.1016/j.physletb.2004.12.006}
  [\eprintth{hep-th/0411225}].
  %%CITATION = doi:10.1016/j.physletb.2004.12.006;%%  

\bibitem{Damour:2007dt}
  T.~Damour, A.~Kleinschmidt and H.~Nicolai,
  ``Constraints and the E10 coset model,''
  \doi{Class.\ Quant.\ Grav.\  {\bf 24} (2007) 6097}{10.1088/0264-9381/24/23/025}
  [\eprintN{0709.2691}{hep-th}].
  %%CITATION = doi:10.1088/0264-9381/24/23/025;%%

\bibitem{Kleinschmidt:2014uwa}
  A.~Kleinschmidt, H.~Nicolai and N.~K.~Chidambaram,
  ``Canonical structure of the E10 model and supersymmetry,''
  \doi{Phys.\ Rev.\ D {\bf 91} (2015) 8,  085039}{10.1103/PhysRevD.91.085039}
  [\eprintN{1411.5893}{hep-th}].
  %%CITATION = doi:10.1103/PhysRevD.91.085039;%%
  
\bibitem{const}  The precise counting of propagating degrees of freedom also hinges on the correct implementation of the constraints~Ref.~\cite{Damour:2007dt}, not all of which are known.   
  
\bibitem{Kleinschmidt:2013eka}
  A.~Kleinschmidt and H.~Nicolai,
  ``On higher spin realizations of $K(E_{10})$,''
  \doi{JHEP {\bf 1308} (2013) 041}{10.1007/JHEP08(2013)041}
  [\eprintN{1307.0413}{hep-th}].
  %%CITATION = doi:10.1007/JHEP08(2013)041;%%  
  
\bibitem{Damour:2014cba}
  T.~Damour and P.~Spindel,
  ``Quantum Supersymmetric Bianchi IX Cosmology,''
  \doi{Phys.\ Rev.\ D {\bf 90} (2014) 10,  103509}{doi:10.1103/PhysRevD.90.103509}
  [arXiv:1406.1309 [gr-qc]].
  %%CITATION = doi:10.1103/PhysRevD.90.103509;%%  
  
\bibitem{Kac} V.~G.~Kac, ``Infinite dimensional Lie algebras,'' Cambridge University Press (1990).  
  
\bibitem{Berman} S.~Berman, 
  ``On generators and relations for certain involutory subalgebras of Kac--Moody Lie algebras,''
   Commun. Algebra {\bf 17} (1989) 3165--3185.  

\bibitem{Damour:2009zc}
  T.~Damour and C.~Hillmann,
  ``Fermionic Kac-Moody Billiards and Supergravity,''
  \doi{JHEP {\bf 0908} (2009) 100}{10.1088/1126-6708/2009/08/100}
  [\eprintN{0906.3116}{hep-th}].
  %%CITATION = doi:10.1088/1126-6708/2009/08/100;%%
 
\bibitem{fn:spin} In the remainder we will not always put quotation marks
 when we talk of `spin'. We trust that readers will understand that this terminology is
 to be taken with a grain of salt, cf. Ref.~\cite{Kleinschmidt:2013eka}.
 
\bibitem{fn:proj} The projectors obey the conditions  $\Pi^{(i)}\Pi^{(j)} = \delta^{ij} \Pi^{(i)}$ and $\sum_j \Pi^{(j)} = {\bf{1}}$. 
 
\bibitem{GHKW} D.~Ghatei, M.~Horn, R.~K\"ohl and S.~Wei\ss,
  ``Spin covers of maximal compact subgroups of Kac-Moody groups and spin-extended Weyl groups,''
  \eprintN{1502.07294}{math.GR}.

\bibitem{ProjNote} The condition employed when fixing these projectors is that they commute with $X(\alpha)$.  

\bibitem{fn:sub} This follows from the fact that $\e$ has infinitely many 
(finite, affine and indefinite) subalgebras that are characterized by their simple root
systems\cite{FN} (which are embedded in the $\e$ root lattice).  The associated subalgebras of $\ke$ are then simply obtained as the associated involutory subalgebras, as in \eqref{xi}.

\bibitem{FN} A.J.~Feingold and H.~Nicolai, 
  ``Subalgebras of hyperbolic Kac-Moody algebras", in:
    Contemp. Math. 343, Amer. Math. Soc, Providence, RI (2004)

\bibitem{Kleinschmidt:2006tm}
  A.~Kleinschmidt and H.~Nicolai,
  ``IIA and IIB spinors from K(E(10)),''
  \doi{Phys.\ Lett.\ B {\bf 637} (2006) 107}{10.1016/j.physletb.2006.04.007}
  [\eprintth{hep-th/0603205}].
  %%CITATION = doi:10.1016/j.physletb.2006.04.007;%%

\bibitem{Jeon:2011vx}
  I.~Jeon, K.~Lee and J.~H.~Park,
  ``Incorporation of fermions into double field theory,''
  \doi{JHEP {\bf 1111} (2011) 025}{10.1007/JHEP11(2011)025}
  [\eprintN{1109.2035}{hep-th}].
  %%CITATION = doi:10.1007/JHEP11(2011)025;%%


\bibitem{Keurentjes:2003yu}
  A.~Keurentjes,
  ``The Topology of U duality (sub)groups,''
  \doi{Class.\ Quant.\ Grav.\  {\bf 21} (2004) 1695}{10.1088/0264-9381/21/6/025}
  [\eprintth{hep-th/0309106}].
  %%CITATION = doi:10.1088/0264-9381/21/6/025;%%

\bibitem{fn:cons} 
  In fact, the consistent decomposability  would
 have to extend to {\em all} the (infinitely many) subgroups of $\KE$, including the ones 
 that descend from affine or indefinite subalgebras of $\e$!

  
\end{thebibliography}
%\printindex[aindx]
%\printindex
\end{document}